\documentclass[aps,prc,10pt,twocolumn]{revtex4-1}
\usepackage{selinput}
\usepackage[T1]{fontenc}
\usepackage{graphicx}
\usepackage{epstopdf}
\usepackage{amsmath,amssymb}
\usepackage{verbatim}
\usepackage[dvipsnames]{xcolor}
\usepackage{geometry}
\usepackage{units}
\usepackage{systeme}
\usepackage{amsmath}
\usepackage[toc,page]{appendix}
\usepackage{float}
\usepackage{subcaption}
\usepackage{lipsum}
\usepackage{caption}
\usepackage{ulem}
\begin{document}
\title{Renormalized critical dynamics and fluctuations in model A}
\author{Nadine Attieh$^{1}$, Nathan Touroux$^{1,2,3}$, Marcus Bluhm$^{1}$, Masakiyo Kitazawa$^{2,4}$, Taklit Sami$^{1}$, Marlene Nahrgang$^{1}$}\email{nadine.attieh@subatech.in2p3.fr}

\affiliation{$^1$SUBATECH UMR 6457 (IMT Atlantique, Universit\'e de Nantes, \\IN2P3/CNRS), 4 rue Alfred Kastler, 44307 Nantes, France}
\affiliation{$^2$Yukawa Institute for Theoretical Physics, Kyoto University, Kyoto 606-8502 Japan}
\affiliation{$^3$Department of Physics, Osaka University, Toyonaka, Osaka 560-0043, Japan}
\affiliation{$^4$J-PARC Branch, KEK Theory Center, Institute of Particle and Nuclear Studies, KEK, 319-1106 Japan}

\newcommand{\MN}[1]{\textcolor{magenta}{#1}}
\newcommand{\NA}[1]{\textcolor{MidnightBlue}{#1}}

\begin{abstract}
In the context of relativistic heavy-ion collisions, we explore the stochastic and dissipative relaxational dynamics of a non-conserved order parameter in a $\lambda\varphi^4$ interaction. The cutoff of the theory is provided by the lattice spacing chosen for our numerical simulations. As a consequence, observables become dependent on that scale. We consider a possible first-order phase transition and an evolution close to a critical point. We demonstrate that using a lattice counterterm restores the expected behavior of the mean, variance and kurtosis: the mean and the variance become lattice spacing independent, and we recover the correct expectation value of the mean, the growth of the variance with the correlation length and the expected minimum in the kurtosis. Our findings hold true in equilibrium and during the dynamical relaxation, and therefore mark an important step towards a fully fluctuating fluid dynamical setup.

\end{abstract}

\maketitle

\section{Introduction}\label{sec:intro}
Dense matter created in relativistic heavy-ion collisions is one of the main subjects of study in ongoing and future experiments. Several theoretical and experimental investigations focus on the QCD temperature-chemical potential phase diagram of this strongly interacting matter~\cite{Luo:2015doi,Bzdak:2019pkr, An:2021wof, Gazdzicki:2020iix, Odyniec:2019kfh, Csanad:2020xbf, Senger:2017oqn}. The equation of state of QCD at zero chemical potential has been known for some years~\cite{Borsanyi:2013bia, HotQCD:2014kol}. In the temperature-baryon chemical potential plane, particular consideration is given to the transition between the deconfined, chirally restored quark-gluon plasma and the hadronic confined phase, where chiral symmetry is broken. Lattice calculations at vanishing baryon chemical potential $\mu_B$ and high temperature $T$, point towards a crossover transition, with a pseudo-critical temperature of $\approx 156$ MeV~\cite{Aoki:2009sc,Fodor:2009ax,Ding:2015ona,HotQCD:2018pds,Borsanyi:2020fev}. At high $\mu_B$ and low $T$, several effective theory calculations predict a first-order transition~\cite{Asakawa:1989bq, Scavenius:2000qd, Antoniou:2002xq}. Questions remain concerning the manifestations of this transition, as well as the location of the conjectured critical point, where extraordinary phenomena associated with long distance correlations and diverging thermodynamic fluctuations are expected to occur~\cite{Stephanov:1998dy}. Because of their dependence on powers of the correlation length which should diverge at the critical point,  moments and quantities that can be derived from them, can act as promising fluctuation observables~\cite{STAR:2020tga,STAR:2014egu,STAR:2010mib}. For example, a characteristic non-monotonic behavior of higher order moments as a function of collision energy can be used to predict proximity to a critical point \cite{Stephanov:2011pb}. However, more thorough theoretical and experimental analyses are needed to confirm that this behavior is indeed evidence of the critical point. In practice, the rapid evolution and short lifetime of the  fireball created during high energy heavy-ion collisions imposes a maximum value on the correlation length, which remains finite, weakening possibles signals of the critical point~\cite{Berdnikov:1999ph, Athanasiou:2010kw}. A dynamical model, with an event-by-event fluctuating initial state coupled to a final hadronic phase where signatures of fluctuations survive, is therefore necessary to study out-of-equilibrium effects at the QCD phase transitions~\cite{Bluhm:2020mpc}. 
\par One such approach consists in coupling the dynamic fluctuations of an order parameter, whose statistical average takes a different value in each phase, to the coarse-grained, macroscopic hydrodynamic evolution of the hot medium. It is well established that hydrodynamics is one of the simplest successful descriptions of the evolution of the fireball produced in relativistic heavy-ion collisions~\cite{Bluhm:2020mpc,Schenke:2010nt}. In such a framework, observables are averaged out, thus reducing the number of degrees of freedom needed to study these systems. However, when describing time-dependent phase transitions, the dynamics of an order parameter needs to be explicitly included~\cite{Mishustin_2014}.
\par So far, two directions of including critical fluctuations in hydrodynamics have been explored. One is deterministic, where the averaged two-point function of the order parameter is coupled to hydrodynamics~\cite{An:2019csj, An:2019osr, Stephanov:2017ghc, Rajagopal:2021doy}. The other is a stochastic approach, where a fluctuating order parameter is explicitly propagated in the dynamically evolving fluid. The latter allows for simpler  implementation in existing event-by-event models, where particles are measured event-by-event and relevant observables can be evaluated in the same manner used experimentally. However, several issues arise in the stochastic approach. The noise introduces an unphysical lattice spacing dependence in numerical calculations and corrections derived from renormalization are inversely proportional to these lattice spacings. Furthermore, noise terms can be a challenge for conventional partial differential equation solvers. Currently, numerically implementing stochastic noise in $3+1$D is significantly complicated and resource consuming (for more details see~\cite{Bluhm:2020mpc} and references therein).
\par The present work is part of the effort to develop the stochastic hydrodynamics approach. While the true dynamical universality class of the QCD critical point in a hydrodynamical environment is argued to be model H~\cite{Son:2004iv} in the Hohenberg-Halperin classification~\cite{Hohenberg:1977ym}, for our purposes here, we focus on the chiral scalar field, which acts as a good order parameter for the chiral phase transition in QCD~\cite{Zinn-Justin:2002ecy, Scavenius:2000qd, Nahrgang:2016eou, Nahrgang:2011sb}. Its stochastic relaxational dynamics was recently numerically investigated in~\cite{Schaefer:2022bfm}. The same field is studied in the framework of non-equilibrium chiral fluid dynamics~\cite{Nahrgang:2011mg}, where its explicit propagation is coupled to the fluid dynamical evolution of the medium created during heavy-ion collisions. We focus here on the renormalization of the relaxational dynamics via the introduction of a lattice counterterm to the effective potential.
\par Numerically solving stochastic equations on a finite lattice requires the introduction of infrared (IR) and ultraviolet (UV) cutoffs. While the IR cutoff reflects the impact of the finite size of the system, the UV cutoff is inversely proportional to the lattice spacing and contributes to the observed unphysical dependence on resolution. Attempts to address this problem include smearing the noise term by a Gaussian distribution~\cite{Murase:2016rhl, Hirano:2018diu}, propagating or coarse graining it over a second grid with larger spacing than usually used in the deterministic case~\cite{Nahrgang:2017oqp, Bluhm:2018plm} or applying a high-mode filter~\cite{Singh:2018dpk}. Though the mean might not be affected, the impact of these procedures on fluctuation observables is unknown and the lattice theory may no longer be equivalent to the continuum theory. In the present work, we look into the possibility of curing this unphysical dependence using lattice renormalization techniques.  Initially, renormalization was developed as a clever mathematical tool to treat infinities that arose in most quantum field theory calculations. However, following developments in renormalization group and effective field theories, these techniques soon came to be regarded as key insights into the underlying physics of phase transitions and critical phenomena~\cite{ZinnJustin,Tetradis:1997bz,Yokota:2016tip,Fu:2022gou}. For a recent study of the relaxational dynamics of a scalar field near a critical point within functional renormalization group approach see~\cite{Batini:2023nan}.
\par As a first step, we focus solely on the stochastic relaxation equation in Eq.~\eqref{eq:relax} governing the dynamical evolution of the order parameter. In Sec.~\ref{sec:relax.model} we describe the model, examine lattice spacing sensitivity and benchmark our numerical methods in the linear approximation. We then describe the lattice spacing dependence in the model with interaction term and discuss lattice renormalization as a possible solution. In Sec.~\ref{sec:mass-CT-on-obs}, we look at the impact of this procedure on the mean and fluctuation observables (as defined in Sec.~\ref{subsec:variables}). We study their behavior before and after renormalization, where a mass counterterm is added to the effective potential~\cite{Cassol-Seewald:2007oak}.  Results are shown in the chirally broken phase and close to the critical point, both at equilibrium and during relaxational evolution.

\section{Relaxational model with no conserved quantities}\label{sec:relax.model}
\subsection{The model}\label{subsec:relax.eq}
We consider a Ginzburg-Landau-Langevin (GLL) type stochastic relaxation equation, Eq.~\eqref{eq:relax}, within the framework of model A in Hohenberg and Halperin's classification~\cite{Hohenberg:1977ym}
\begin{equation}
    \frac{\partial^2\varphi}{\partial t^2} -\nabla^2\varphi +\eta\frac{\partial \varphi}{\partial t} + \frac{\partial V_{\rm eff}[\varphi]}{\partial\varphi} = \xi \,,
    \label{eq:relax}
\end{equation}
which describes the non-equilibrium evolution of the scalar field $\varphi=\varphi(\vec{x}, t)$. The damping coefficient $\eta$ imposes the time scale of the system and encodes the intensity of the dissipation. $\xi=\xi(\vec{x}, t)$ is a white thermal noise, i.e. local in space and time. It is entirely characterized by its mean and variance
\begin{subequations}
    \begin{equation}
        \langle\xi(\vec{x},t)\rangle=0 \,,
        \label{eq:xi.mean}
    \end{equation}
    \begin{equation}
        \langle\xi(\vec{x},t)\xi(\vec{x}',t')\rangle=2\eta T\ \delta(\vec{x}-\vec{x}')\delta(t-t') \,.
        \label{eq:xi.var}
    \end{equation}
\end{subequations}
This choice ensures that in the continuum limit $\varphi(\vec{x}, t)$ relaxes to the correct equilibrium value, guarantees a proper equilibrium distribution and satisfies the fluctuation-dissipation theorem. In principle, the damping coefficient and the noise correlator can be calculated from 2PI (two-particle irreducible) effective action approaches, see e.g.~\cite{Nahrgang:2011mg}. However in the present work the focus lies on the numerical implementation. We therefore take $\eta$ as a parameter.
\par $V_\text{eff}[\varphi]$ is a Ginzburg-Landau quartic effective potential with phase transition
\begin{equation}
    V_{\rm eff}(\varphi)=\frac{1}{2}\epsilon\varphi^2+\frac{1}{4}\lambda\varphi^4\,,
 \label{eq:Veff}
\end{equation}
where $\lambda$ is the coupling coefficient and $\epsilon$ is the square mass parameter, which encodes the phase transition. $\epsilon<0$ describes a first order transition, the system is in the chirally broken phase and the potential has a double well structure. Positive small values of $\epsilon$, describe the system in the vicinity of the critical point and the single well potential associated with $\epsilon>0$ is flattened out.

\subsection{Gauss approximation $\lambda=0$ and benchmarking}\label{subsec:gauss}
We wish to benchmark our numerical framework against known analytical solutions. Solving Eq.~\eqref{eq:relax} analytically is only possible under very specific conditions, such as the linear approximation of the effective potential in Eq.~\eqref{eq:Veff} with $\epsilon>0$. We take $\lambda=0$, and derive the two-point equilibrium correlation function in three spatial dimensions as a function of $r=|\Vec{x}-\Vec{x}'|$~\cite{Landau:1980mil}
\begin{equation}
    C(\vec{r})=\frac{T}{4\pi r}e^{-\frac{r}{r_c}}\,,
    \label{eq:gauss.corr}
\end{equation}
where $r_c=\sqrt{\frac{1}{\epsilon}}$ is the correlation length. This correlation function diverges for $r=0$.
\par In the following, we choose units such that all quantities are dimensionless (see Appendix~\ref{app:dim-less}). For the numerical solution, we consider a static cubic system with sides $L=20$ at fixed temperature, with $N$ cells in each direction. We use periodic boundary conditions. We take $dx=dy=dz=\nicefrac{L}{N}$ and refer to the lattice spacing simply as $dx$. We vary the lattice spacing by changing $N$, corresponding to $dx=0.625,\ 0.417,\ 0.3125,\ 0.25$ and $0.208$. Because we are interested in studying the correlation function in equilibrium, we wait until the system reaches equilibrium at around $t_{\rm fin}=60$, and evaluate $C(\vec{r})=\langle \varphi(0,t) \varphi(\vec{r},t) \rangle$. The time step is chosen as $dt=0.1$ and we set $T=\eta=1$. The number of noise configurations is set to $10^4$.
\par
\begin{figure}[ht]
    \centering
    \includegraphics[width=0.485\textwidth]{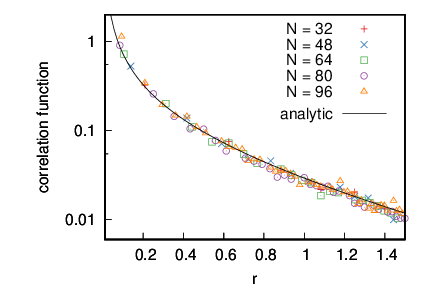}
    \caption{Numerical results for different lattice spacings $dx=\nicefrac{20}{N}$ reproduce the expected analytical correlation function in the Gauss approximation, Eq.~\eqref{eq:gauss.corr}, with $T=1$, $\epsilon=1$ and $r_c=1$.}
    \label{fig:gauss.corr}
\end{figure}

 As expected, in the absence of interaction terms, the equilibrium correlation function shows no lattice spacing dependence for finite distances (see Fig.~\ref{fig:gauss.corr}), and the results are in agreement with the analytical expression in Eq.~\eqref{eq:gauss.corr}.
\begin{figure}[ht]
    \centering
    \includegraphics[width=0.49\textwidth]{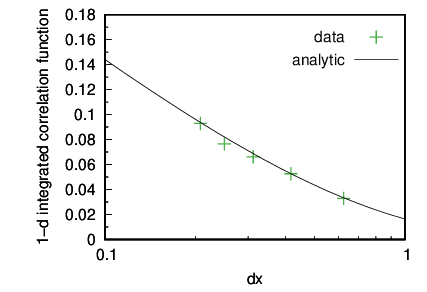}
    \caption{Numerical results compared to the expected logarithmic lattice spacing dependence of the one-dimensional integral Eq.~\eqref{eq:int.corr.gauss} of the correlation function for $X=3$.}
    \label{fig:gauss.corr.int}
\end{figure}
\par The three-dimensional integral of $C(\vec{r})$ in Eq.~\eqref{eq:gauss.corr} is also lattice spacing independent. In order to better visualize the lattice spacing dependence, we look into the one-dimensional integral of the correlation function along the radial axis as a purely mathematical quantity to check the agreement between numerical data and the analytically expected logarithmic lattice spacing dependence. Expanding $e^{-\nicefrac{r}{r_c}}$ for small $r$ in Eq.~\eqref{eq:gauss.corr} we find
\begin{equation}
    \int_{dx}^X \frac{T}{4\pi r}e^{-\frac{r}{r_c}}\ dr\propto \ln(X)-\ln(dx)\,,
    \label{eq:int.corr.gauss}
\end{equation}
where $X$ is an arbitrary upper integration limit, constrained by the size of the system. Taking $X=3$, we show in Fig.~\ref{fig:gauss.corr.int} the comparison between numerical results and expected logarithmic dependence of the one-dimensional integral of $C(\vec{r})$ on the lattice spacing. 
\par Having successfully reproduced the analytical results and benchmarked our numerical framework, we now introduce renormalization and apply it to the system with the full potential of Eq.~\eqref{eq:Veff}.

\subsection{Lattice renormalization for $\lambda\neq 0$}\label{subsec:lattice.renorm}
Numerically solving the stochastic relaxation equation Eq.~\eqref{eq:relax}, using the full effective potential with $\lambda\neq 0$ in Eq.~\eqref{eq:Veff}, introduces a lattice spacing dependence due to the need for a UV-cutoff. Lattice renormalization is increasingly used in perturbative quantum field theory to treat this sensitivity. For the Ginzburg-Landau quartic potential only mass renormalization is required. Divergent diagrams in the perturbative expansion were identified and their contribution can be compensated by a quadratic mass counterterm, see~\cite{Cassol-Seewald:2007oak, Gagne:1999nh, Farakos_1994, Farakos:1994xh}. The impact of this renormalization on the mean in particular was studied in~\cite{Cassol-Seewald:2007oak}. The authors considered a similar stochastic GLL type equation and restored lattice spacing independence of the mean in equilibrium for a system in the chirally broken phase. In our work, we are particularly interested in the dynamics of fluctuation observables. In addition, we examine the behavior of these fluctuations in two distinct situations: the same conditions of broken symmetry as explored in~\cite{Cassol-Seewald:2007oak}, and closer to the critical point. In the latter case, the perturbative expansion is expected to break down. We aim to explore how close to criticality renormalization procedures can still be useful.
\par The correction to the effective potential following the inclusion of the counterterm derived from mass renormalization is given by 
\begin{widetext}
\begin{equation} 
    V_{\rm CT}=\Bigg\{-\frac{3\lambda\Sigma}{4\pi}\frac{T}{{\rm d}x} \\
    +\frac{3}{8}\left(\frac{\lambda T}{\pi}\right)^2\left[\ln\left(\frac{6}{{M\rm d}x}\right)+\zeta\right]\Bigg\}\frac{\varphi^2}{2}\,,
    \label{eq:CT}
\end{equation}
\end{widetext}
where $M$ is the renormalization scale. Its contribution is logarithmic and therefore weak. $\Sigma$ and $\zeta$ are constants that appear in the renormalization procedure:
\begin{equation*}
    \Sigma\approx3.1759\quad ;\quad \zeta\approx0.09\,.
\end{equation*}
In the following, we explicitly show the $dx$ dependence of the bare model, then explore the effects of adding the mass counterterm Eq.~\eqref{eq:CT} to renormalize the potential in Eq.~\eqref{eq:Veff}.

\section{Impact of renormalization on mean and fluctuation observables}\label{sec:mass-CT-on-obs}
Solving Eq.~\eqref{eq:relax} and extracting fluctuation observables in three spatial dimensions with non-linear terms (with or without the counterterm), requires extensive numerical simulations. We discretize Eq.~\eqref{eq:relax} using a standard leapfrog scheme and apply periodic boundary conditions on the cubic lattice of sides $L=20$, described in Sec.~\ref{subsec:gauss}. We integrate the field over the volume of a sphere, starting at the center of the cube (radius $=0$) and expanding to its sides (radius $\leq\nicefrac{L}{2}$). Again, units are chosen to make all quantities dimensionless. We take $T=M=\eta=1$. $\lambda$, dimensionless by nature (see appendix~\ref{app:dim-less}), is set to $0.25$. The number of noise configurations is again set to $10^4$.
\par For more time efficient use of resources, the numerical evaluation of the dynamical field $\varphi(\Vec{x}, t)$ and subsequent computation of relevant quantities, such as the correlation function and the different moments, are performed on GPU.

\subsection{Mean and fluctuation observables}\label{subsec:variables}
We start by defining the observables of interest. The volume average of the non-conserved order parameter is given by
\begin{equation}
    \varphi_V(t)=\frac{1}{V}\int_V d^3 x\,\varphi(\Vec{x}, t) \,.
    \label{eq:mean}
\end{equation}
$\varphi_V(t)$ is evaluated for each noise configuration. We thus obtain a distribution of $\varphi_V(t)$ over the number of noise configurations. From this distribution, we take the mean
\begin{equation}
    \Phi = \langle\varphi_V(t)\rangle_{\rm conf} \,,
    \label{eq:mean1}
\end{equation}
and the higher-order cumulants of the fluctuations. These quantities can then be studied at late times when the system governed by Eq.~\eqref{eq:relax} has relaxed to equilibrium. The second-order cumulant corresponds to the variance 
\begin{equation}
    \sigma^2= \langle\left(\varphi_V(t)-\Phi\right)^2\rangle_{\rm conf} \,.
    \label{eq:var}
\end{equation}
In this definition, the variance has a trivial volume dependence, which we scale out in the results showing $\sigma^2\times V$. 
As pointed out in Sec.~\ref{sec:intro}, higher order cumulants are thought to be more indicative of the closeness to the critical point than quadratic moments. The experimentally relevant quantity is $\kappa\sigma^2$, which corresponds to the ratio of the fourth-order cumulant over the variance. Here
\begin{equation}
    \kappa= \frac{\langle\left(\varphi_V(t)-\Phi\right)^4\rangle_{\rm conf}}{\sigma^4}\,,
    \label{eq:kappa}
\end{equation}
refers to the volume-independent kurtosis. As a result, we scale out the trivial volume dependence of the variance again and show $\kappa\sigma^2\times V$. As the system approaches criticality, the probability distribution of $\varphi_V(t)$ is expected to become non-Gaussian. The kurtosis then takes a finite non-zero value~\cite{Bzdak:2019pkr,Stephanov:2011pb}.
\par The fluctuation observables are studied in our numerical calculations for three cases:
\begin{itemize}
    \item $\epsilon=-1$, chiral symmetry is broken and we have a double well potential. Here, we investigate the mean $\Phi$ and the variance $\sigma^2$ at equilibrium and during relaxation;
    \item $\epsilon=0.1$, indicating proximity to a critical point where we see a flat, single well potential. We look at the same two observables as before, again at equilibrium and during relaxation;
    \item $\epsilon=0.01$, the system is even closer to the critical point and the single well potential becomes flatter. Here, we are interested in the dynamical evolution of the kurtosis as $\kappa\sigma^2$.
\end{itemize}
In the following we study the equilibrium values at a certain time $t_{\text{fin}}$, after which all relevant quantities do not change anymore in time. In this case, we vary the volume of integration by changing the radius $X$. For the dynamical evolution the radius is fixed at $X=8$, and the results are shown between $t=0$ and variable $t=t_{\text{fin}}$.

\subsection{Mean and variance at equilibrium}\label{subsec:eq}
In order to study the equilibrium physics, we let the system evolve until $\Phi$ and $\sigma^2$ reach stable values over time.
\begin{figure*}
\centering
    \begin{subfigure}{0.49\textwidth}
    \includegraphics[width=\textwidth]{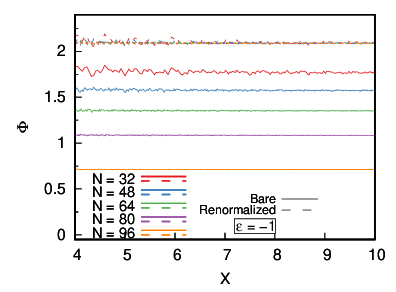}
    \label{subfig:eps-1.mean-eq}
\end{subfigure}
\begin{subfigure}{0.49\textwidth}
    \includegraphics[width=\textwidth]{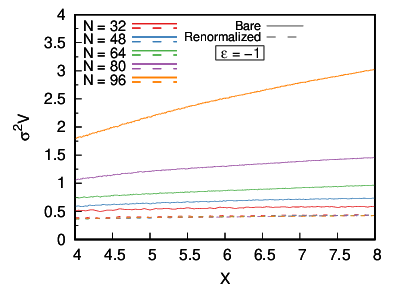}
    \label{subfig:eps-1.var-eq}
\end{subfigure}
\caption{Mean (left) and variance (right) at equilibrium in the chirally broken phase for $\epsilon=-1$ (first-order phase transition situation) for different lattice spacings $dx=\nicefrac{20}{N}$. Dashed (solid) lines are with (without) counterterm.}
\label{fig:eps-1.eq}
\end{figure*}
\begin{figure}
    \centering
    \includegraphics[width=0.49\textwidth]{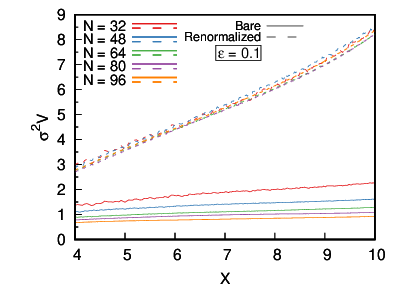}
    \caption{Variance at equilibrium close to the critical point for $\epsilon=0.1$ and different lattice spacings $dx=\nicefrac{20}{N}$. Dashed (solid) lines are with (without) counterterm.}
    \label{fig:eps0.1-eq}
\end{figure}
For $\epsilon=-1$ the system is in the chirally broken phase. Results are shown in Fig.~\ref{fig:eps-1.eq}. The left panel corresponds to the mean $\Phi$, as defined in Eq.~\eqref{eq:mean1}, while the left panel shows the variance as $\sigma^2 \times V$, with $\sigma^2$ defined in Eq.~\eqref{eq:var}. Both quantities are shown as functions of the volume, represented by the radius $X$. The solid lines correspond to the bare system results without counterterm for different cell numbers $N$. Here, we observe a strong dependence of the equilibrium quantities on $dx$ over the full range of $X$. The dashed lines show the same quantities for the renormalized system. Curves for different lattice spacings lie perfectly on top of each other. The unphysical lattice spacing dependence has been cured. Furthermore, in the case of the mean, the renormalized value is larger than in the bare case and corresponds to the correct minimum of the potential for the selected values of $\epsilon$ and $\lambda$. The restoration of the lattice spacing independence of the mean has already been shown in \cite{Cassol-Seewald:2007oak} for a volume corresponding to the entire lattice. We confirm that this holds for all subvolumes $V$. 
\par In our study, we go beyond the mean and demonstrate the restoration of the lattice spacing independence for the variance as well, as shown on the right-hand side of Fig.~\ref{fig:eps-1.eq}. The solid lines for the bare results exhibit the same pronounced $dx$-dependence as the mean. When the system is renormalized, curves in dashed lines for different lattice spacings merge together over the full $X$ range. A false rising trend in the variance for the smallest $dx$ is also corrected.
\par The case where the system is close to criticality, i.e.~for $\epsilon=0.1$, was not addressed in~\cite{Cassol-Seewald:2007oak}. Our results are shown in Fig.~\ref{fig:eps0.1-eq}. We only consider the variance, represented as $\sigma^2 \times V$, at equilibrium since the mean vanishes and is therefore of little interest. In the vicinity of the critical point, renormalization (dashed lines) also cures the lattice spacing dependence observed in the bare case (solid lines), over the considered volumes. As expected, since the variance scales as some positive power of the correlation length, which theoretically diverges close to the critical point, long-range fluctuations add up and the variance increases with the volume. This expected trend is seen only after including the counterterm. The renormalized variance is also significantly larger than the bare one.

\subsection{Dynamical relaxation of the mean and the variance}\label{subsec:dyn}
We now look at the relaxational evolution of both observables on a sphere of fixed radius $X=8$ between $t=0$ and $t_{\text{fin}}=60$. While the initial state was of no importance for the long-time, equilibrium behavior of the system, the dynamical relaxation process itself depends on the initial conditions. For the present simulations, the fields are initialized homogeneously such that $\varphi(\vec{x},t=0) = \varphi_0$. A constant initial value $\varphi_0$ implies that the variance at $t=0$ is zero. This imposes a strong non-equilibrium constraint, especially in the vicinity of the critical point.
\begin{figure*}
\centering
    \begin{subfigure}{0.49\textwidth}
    \includegraphics[width=\textwidth]{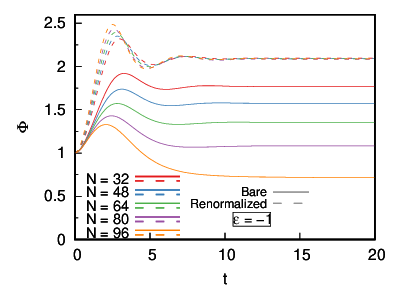}
    \label{subfig:eps-1.mean-time.evol}
\end{subfigure}
\begin{subfigure}{0.49\textwidth}
    \includegraphics[width=\textwidth]{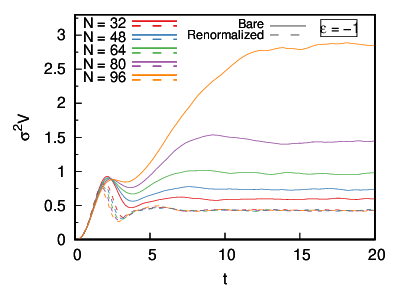}
    \label{subfig:eps-1.var-time.evol}
\end{subfigure}
\caption{Dynamical evolution of the mean (left) and the variance (right) in the chirally broken phase with $\epsilon=-1$ and for different lattice spacings $dx=\nicefrac{20}{N}$. Dashed (solid) lines are with (without) counterterm.}
\label{fig:eps-1.time-evol}
\end{figure*}
\par For $\epsilon=-1$, we choose as initial value $\varphi_0=1$. This choice avoids large peaks in $\sigma^2$ as the system starts to relax towards equilibrium. Results for different $dx$-values are depicted in Fig.~\ref{fig:eps-1.time-evol}: the mean $\Phi$ is again shown on the left, while the variance as $\sigma^2 \times V$ is shown on the right as functions of time. We only show results up to $t=20$, where equilibrium is already established and no further relevant information can be gained afterwards. The solid lines correspond to the bare system where the expected lattice spacing dependence is observed. The dashed lines correspond to the same quantities following renormalization, where the $dx$-dependence is significantly reduced in both quantities.
\begin{figure*}
\centering
    \begin{subfigure}{0.49\textwidth}
    \includegraphics[width=\textwidth]{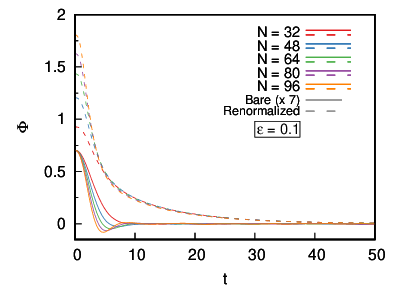}
    \label{subfig:eps0.1-mean-time.evol}
\end{subfigure}
\begin{subfigure}{0.49\textwidth}
    \includegraphics[width=\textwidth]{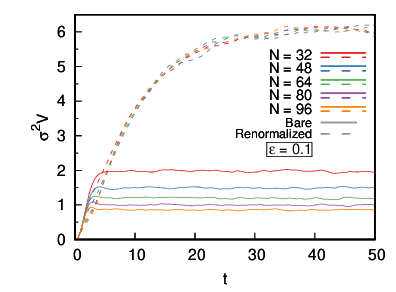}
    \label{subfig:eps0.1-var-time.evol}
\end{subfigure}
\caption{Dynamical evolution of the mean (left) and the variance (right) close to critical point for $\epsilon=0.1$ and for different lattice spacings $dx=\nicefrac{20}{N}$. Dashed (solid) lines are with (without) counterterm. The mean without counterterm is scaled by a factor of $7$ for a better visualization.}
\label{fig:eps0.1-time.evol}
\end{figure*}
\par In the case of proximity to the critical point, expressed by $\epsilon=0.1$, we encounter the following delicate situation: due to the homogeneous initial condition the fluctuations, and thus the fluctuation-induced lattice spacing sensitivity, need some time to develop. The addition of the $dx$-dependent counterterm in Eq.~\eqref{eq:CT} to the effective potential artificially introduces a lattice spacing sensitivity at early times. The counterterm shifts the positive bare mass term $\epsilon=0.1$ towards $dx$-dependent negative dressed values. As a consequence, the effective potential initially resembles a double well potential with a non-zero minimum, rather than a single well with a single minimum at $\varphi=0$. For small fixed initial value the restoring force is different for different $dx$, and the initial evolution would move the field towards the $dx$-dependent positive minimum. To avoid this artefact we apply the following procedure: we chose the initial value of the homogeneous field for the bare system as $\varphi_0=0.1$. In order to draw a meaningful comparison with the renormalized system, we adapt the initial value $\varphi_0(dx)$ for this case as a function of $dx$. We determine $\varphi_0(dx)$ such that the initial restoring force is the same as for the bare system 
\begin{equation*}
    \frac{\partial({V_{\rm eff}})^{\rm ren}}{\partial\varphi}\Bigg|_{\varphi_0=\varphi_0(dx)}=\frac{\partial({V_{\rm eff}})^{\rm bare}}{\partial\varphi}\Bigg|_{\varphi_0=0.1} \,.
\end{equation*}
In Fig.~\ref{fig:eps0.1-time.evol}, we show the mean (left) and the variance as $\sigma^2 \times V$ (right) obtained with this procedure as a function of time for the different values of $dx$. We note that the relaxation time is longer compared to the $\epsilon=-1$ case, especially after renormalization. Following the addition of the counterterm, $\varphi=0$ is no longer initially the minimum. This creates a barrier, delaying the relaxation towards the correct equilibrium value until the fluctuations kick in and restore the flattened single well effective potential. Results are therefore shown up to $t=50$. The lattice spacing dependence seen in the bare system (solid lines) is eliminated by the addition of the counterterm (dashed lines). As discussed above, a quirk remains in the case of the mean: initially, in the absence of fluctuations, the chosen values for $\varphi_0(dx)$ differ according to $dx$. Therefore at early times, the curves for the different $dx$ start at different values. Afterwards, the fluctuations start to relax to equilibrium, as seen for the variance on the right plot of Fig. \ref{fig:eps0.1-time.evol} and the evolution becomes lattice spacing independent. Surprisingly, the evolution of the mean becomes $dx$ independent at around $t=5$, long before the variance has reached its equilibrium value, see Fig.~\ref{fig:eps0.1-time.evol}. This indicates that the equilibrium counterterm we are using, Eq.~\eqref{eq:CT},  can be applied even in cases where fluctuations are not fully equilibrated. Such a situation typically arises near a critical point in a dynamical environment due to critical slowing down.

\subsection{Dynamical relaxation of the kurtosis}\label{subsec:kurt}
In this section, we investigate the relaxational dynamics of $\kappa\sigma^2$ before and after including the counterterm. In the same conditions as above for $\epsilon=0.1$, $\kappa\sigma^2$ shows a slight shift towards negative values in the renormalized case. However, a solid conclusion cannot be drawn without significantly more statistics. An increase of several orders of magnitude seems to be necessary. In order to get better insight into the impact of renormalization on $\kappa\sigma^2$, we look at the system even closer to the critical point and take $\epsilon=0.01$ instead.
\par Interest in higher-order cumulants as indicators of criticality is partly due to the fact that they scale with larger powers of the correlation length and may change signs~\cite{Stephanov:2008qz,Asakawa:2009aj}. Because of the first feature, their relaxation towards equilibrium may require significantly longer time \cite{Mukherjee:2015swa}. For $\epsilon=0.01$, we verified that allowing the bare system to evolve up to $t_{\rm fin}=60$ is sufficient for the second- and fourth-order cumulants to reach equilibrium values. However, following renormalization, we find that this time needs to be increased to $t_{\rm fin}=300$, to ensure complete equilibration of the second- and fourth-order cumulants. Because of the increased demand in time and energy consumption, and in light of sustainability concerns (see Appendix \ref{app:CO2}), we limit our study in this case to four values of $dx$, corresponding to $N=32, 48, 64$ and $80$.
\par
\begin{figure}
    \centering
    \includegraphics[width=0.49\textwidth]{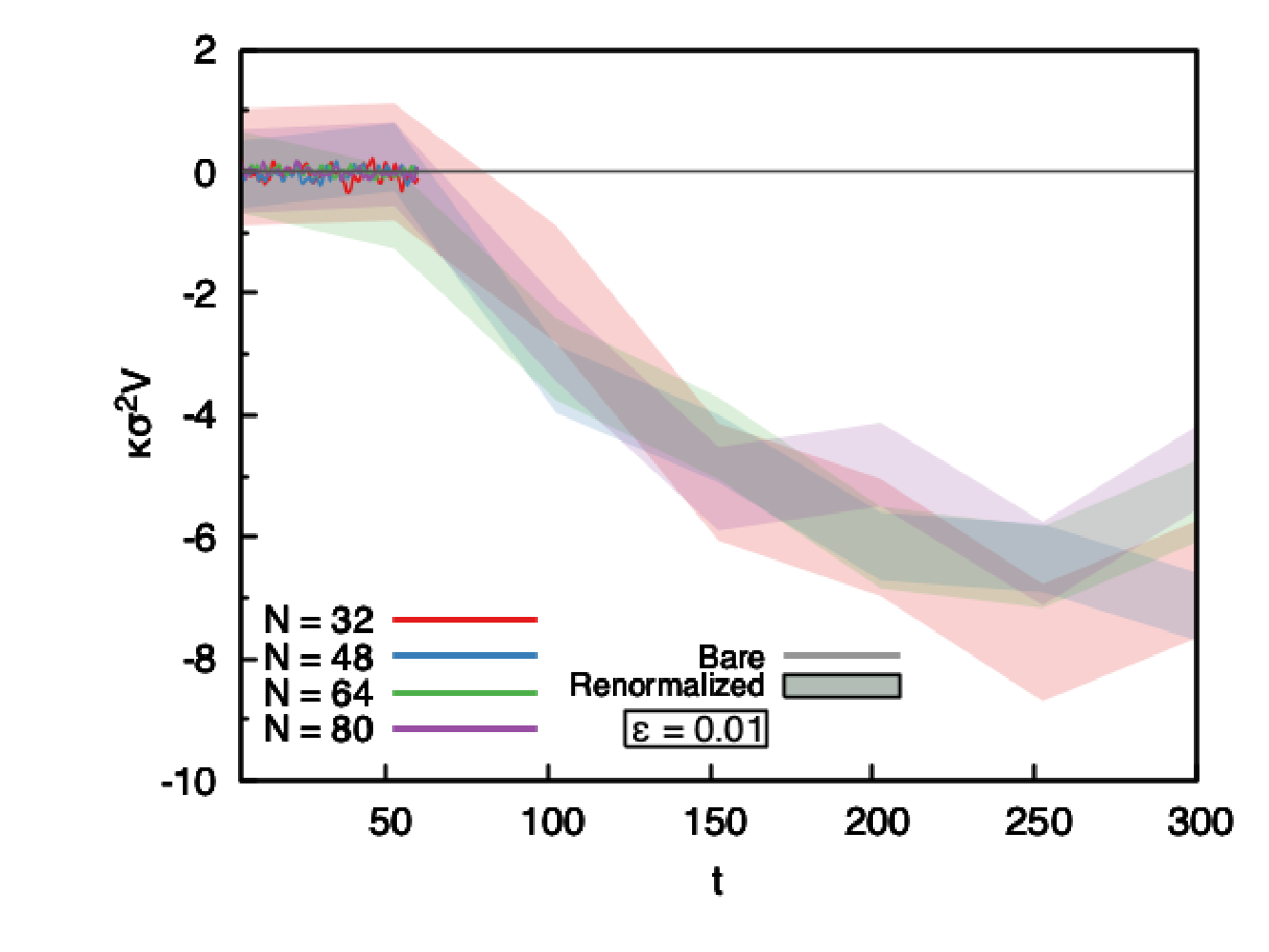}
    \caption{Dynamical evolution of $\kappa\sigma^2$ for $\epsilon=0.01$. Results before addition of the counterterm are shown as solid lines up to $t_{\text{fin}}=60$. The results following renormalization are shown as shaded areas over the complete time range.}
    \label{fig:ksig2-eps0.01}
\end{figure}

 Due to limited statistics we present the renormalized results as shaded bands in Fig.~\ref{fig:ksig2-eps0.01}. The bands are obtained using the following procedure: for each value of $dx$, a standard deviation is evaluated over all timesteps after the relevant cumulants are equilibrated, i.e. between $t=200$ and $t_{\rm fin}=300$. Then, all the data points are grouped in successive bins of 100 timesteps in size. For each bin the average is determined and placed in the middle of this bin. The above standard deviation is then applied to each bin over the full time evolution, i.e. between $t=0$ and $t_{\rm fin}=300$ giving rise to the shaded areas depicted in Fig.~\ref{fig:ksig2-eps0.01}.
\par As discussed in Sec.~\ref{subsec:variables}, when the system approaches the critical point, the probability distribution loses its Gaussian nature. The behavior of higher-order cumulants and their derived quantities is particularly well suited as indicator of this feature. For example, the kurtosis and consequently $\kappa\sigma^2$ are expected to have finite non-zero equilibrium values close to the critical point. In Fig.~\ref{fig:ksig2-eps0.01} results for the bare system (solid lines), where $\kappa\sigma^2$ is essentially zero, are shown up to $t_{\rm fin}=60$. This behavior will not change beyond that time (solid grey line) as the cumulants have individually equilibrated already. However, Fig.~\ref{fig:ksig2-eps0.01} also shows that lattice renormalization (shaded areas) allows the recovery of the expected non-Gaussian behavior in close proximity to the critical point: $\kappa\sigma^2$ deviates from zero as the individual cumulants equilibrate. Lastly, we note that no robust conclusions can be reached regarding an improvement of lattice spacing dependence within the available statistics.

\section{Conclusion}\label{sec:conclusion}
Lattice spacing dependence is one of the more complex issues impacting studies of heavy-ion collisions in stochastic hydrodynamics. On an event-by-event basis this problem still needs to be thoroughly addressed with renormalization techniques. In this work, we looked at lattice renormalization as a possible solution in the case of a fluctuating order parameter whose dynamics is governed by a stochastic relaxation equation.
\par First, we successfully benchmarked our numerical framework against known analytical results in the linear approximation: the two-point equilibrium correlation function and its one-dimensional integral. Then in the full model, we defined the mean and fluctuation observables and introduced a counterterm to renormalize the mass term.
\par For the nonlinear and bare stochastic model we observe the expected lattice spacing dependence. Our numerical results demonstrate that by introducing a lattice counterterm to the effective potential this lattice spacing dependence is cured. This had so far been studied only for the mean in the double well potential at equilibrium~\cite{Cassol-Seewald:2007oak}. We can now show that this is also true for the mean and the variance in a double well potential and close to a critical point.
\par The renormalized variance in equilibrium shows an increasing trend with the volume near a critical point in accordance with a large correlation length. The dynamical relaxation process is impacted by our constant initial conditions, but lattice spacing dependence at early times is cured as the fluctuations in the system develop. Surprisingly, the existence of fluctuations seems sufficient as lattice spacing independence is achieved long before the variance equilibrates. This can be an important feature for future studies of the critical point in dynamical simulations of heavy-ion collisions. Here, due to critical slowing down the fluctuations are expected to be off-equilibrium.
\par Next, we looked at the kurtosis. Due to the small effect of renormalization close to a critical point with $\epsilon = 0.1$ and the huge amount of noise configurations that would be necessary for a quantitative statement, we moved even closer to the critical point by choosing $\epsilon=0.01$. We see that the expected negative values for $\kappa\sigma^2$ are recovered only after including the lattice counterterm. The bare results show no signs of non-Gaussianity.
\par The presented procedure is to be implemented in the broader context of developing a framework of renormalized stochastic hydrodynamics. In future work it will be interesting to couple the renormalized chiral dynamics presented here to a hydrodynamical evolution, along the lines of nonequilibrium chiral fluid dynamics~\cite{Nahrgang:2011mg,Nahrgang:2011vn,Herold:2017day}. In addition, it will be useful to study the diffusive dynamics of a conserved order parameter, as in model B of Hohenberg and Halperin~\cite{Hohenberg:1977ym}, or its coupling to a hydrodynamical environment as in model H, recently studied in~\cite{Chattopadhyay:2024jlh}. Ultimately, the goal is to apply the renormalization of the full stochastic hydrodynamics to heavy-ion collisions near a critical point.
\par The fluctuation observables tested in this work can be extracted for particle multiplicities from experiments aiming to probe the QCD phase diagram with heavy-ion collisions. If confirmed in increasingly realistic stochastic models of hydrodynamics, lattice renormalization could produce robust results that make reliable predictions possible for experimental studies.

\begin{appendices}

\section{Dimensionless quantities}\label{app:dim-less}
In a symmetry broken Ginzburg-Landau quartic potential~\cite{Cassol-Seewald:2007oak}, the square mass parameter is defined as
\begin{equation*}
    \epsilon=\frac{\lambda}{4}(T^2-T_c^2)\,.
\end{equation*}
In order to make all quantities dimensionless in our analysis, we express them in units of the critical temperature $T_c$. We can then write
\begin{align*}
    \varphi\rightarrow&\frac{\varphi}{T_c}\,,\\
    \epsilon\rightarrow&\frac{\epsilon}{T_c^2}=\frac{\lambda}{4}\Big (\frac{T^2}{T_c^2}-1\Big )\,,\\
    T,\ \eta,\ M\rightarrow&\frac{T}{T_c},\ \frac{\eta}{T_c},\ \frac{M}{T_c}\,,\\
    dx,\ dt\rightarrow&T_c dx,\ T_c dt\,,\\
    V_{\rm eff}\rightarrow&\frac{V_{\rm eff}}{T_c^4}\,.
\end{align*}
Equations~\eqref{eq:relax} and~\eqref{eq:Veff} are homogeneous in the chosen unit system when $\lambda$ is dimensionless.

\section{Carbon footprint}\label{app:CO2}
In a global context of major climate and environmental concerns, as well as social and economical issues, we try to get some insight into the carbon footprint of this work. For increased speed and computing power, numerical calculations were carried out on GPUs at the National Institute of Nuclear and Particle Physics (IN2P3) computing center in Villeurbanne, France.
\par We estimate that roughly 87 kGPUh (Thermal Design Power or $TDP=250 -300$ W~\cite{Nvidia}), and 1200 KCPUh ($TDP_\text{max}=8.5$ W per core~\cite{Intel}), were required to obtain the final results shown in this work, as well as tests, trial and error runs, general managing functions, data analysis and formatting. Average CO$_2$ emissions in France for 2023 were at $53$ gCO$_2$eq/kWh~\cite{CO2}. Using respective TDPs, a cautious estimate of $\approx 1.8$ tCO$_2$eq were emitted over the course of several months. $1$ tCO$_2$eq are emitted for one passenger on a round trip flight Paris-New York or for an average car over a distance of 5000 km~\cite{RTE}.
\par TDP, used for simplicity and accessibility reasons, is generally regarded as a poor surrogate for actual power usage, giving order-of-magnitude results~\cite{osti_1838264}. In practice, clock-frequency, processing unit efficiency and other power-saving measures~\cite{Markin:2023fxx} need to be taken into account for more accurate estimates.
\par Finally it is worth noting that because of France's energy policy, relying largely on nuclear power~\cite{RTE}, CO$_2$ emissions are considerably lower than in other countries, where energy production is more fossil fuel dependent. For example, in 2023, average $CO_2$ emissions reached $\approx 411$ gCO$_2$eq/kWh in the United States and $\approx 400$ gCO$_2$eq/kWh in Germany~\cite{CO2}. The same calculations would have roughly resulted in emissions of $\approx 14$ tCO$_2$eq and $\approx 13$ tCO$_2$eq respectively in each country.

\end{appendices}

\bibliography{biblio.bib}

\end{document}